\documentclass[12pt]{article}
\usepackage{authblk}
\usepackage[bookmarksnumbered, colorlinks, plainpages]{hyperref}
\usepackage{amsmath, amsthm, amscd, amsfonts, amssymb, graphicx, color, booktabs}
\numberwithin{equation}{section}
\definecolor{email}{rgb}{0.00,0.00,0.84}
\begin{document}
\setcounter{page}{1}

\title{\large \bf 12th Workshop on the CKM Unitarity Triangle\\ Santiago de Compostela, 18-22 September 2023 \\ \vspace{0.3cm}
\LARGE tt+X measurements by CMS and ATLAS}

\author{B\'arbara \'Alvarez Gonz\'alez$^{1}$ on behalf of the CMS and ATLAS Collaborations\footnote{"Copyright 2024 CERN for the benefit of the ATLAS and CMS Collaborations. Reproduction of this article or parts of it is allowed as specified in the CC-BY-4.0 license" }}
\affil{$^{1}$Universidad de Oviedo, Instituto Universitario de Ciencias y Tecnolog\'ias Espaciales de
Asturias (ICTEA), Oviedo, Spain}
\maketitle

\begin{abstract}
These proceedings present the observation of the 4-top production process and the latest results of the production of top-quark pairs associated with W and $\gamma$ bosons ($t\bar{t}W$ and $t\bar{t}\gamma$) and b jets ($t\bar{t}b\bar{b}$) at a collision energy of 13~TeV carried out by the CMS and ATLAS Collaborations.
\end{abstract} \maketitle

\section{Introduction}
The top quark is a unique particle. It is the most massive of all observed elementary particles and the only quark that decays before hadronisation, which allows direct access to its properties. Top quark properties are studied extensively in the CMS~\cite{CMS} and ATLAS~\cite{ATLAS} Collaborations.
These studies provide important input to test theoretical calculations and have the potential to reveal deviations from the standard model (SM) predictions. The large amount of LHC data recorded during Run 2 allows probing SM processes with small production cross sections which become fully accessible inclusively and differentially.

These proceedings present the observation of 4-top production with more than 5$\sigma$ significance, the inclusive and first differential $t\bar{t}W$ cross sections, the most precise measurements of $t\bar{t}b\bar{b}$ production up to date and the first measurement of the charge asymmetry of top-quark pairs in production in $t\bar{t}\gamma$.

\section{Observation of the 4-top production process}
\begin{figure}[h]
    \centering
    \includegraphics[width=0.85\linewidth]{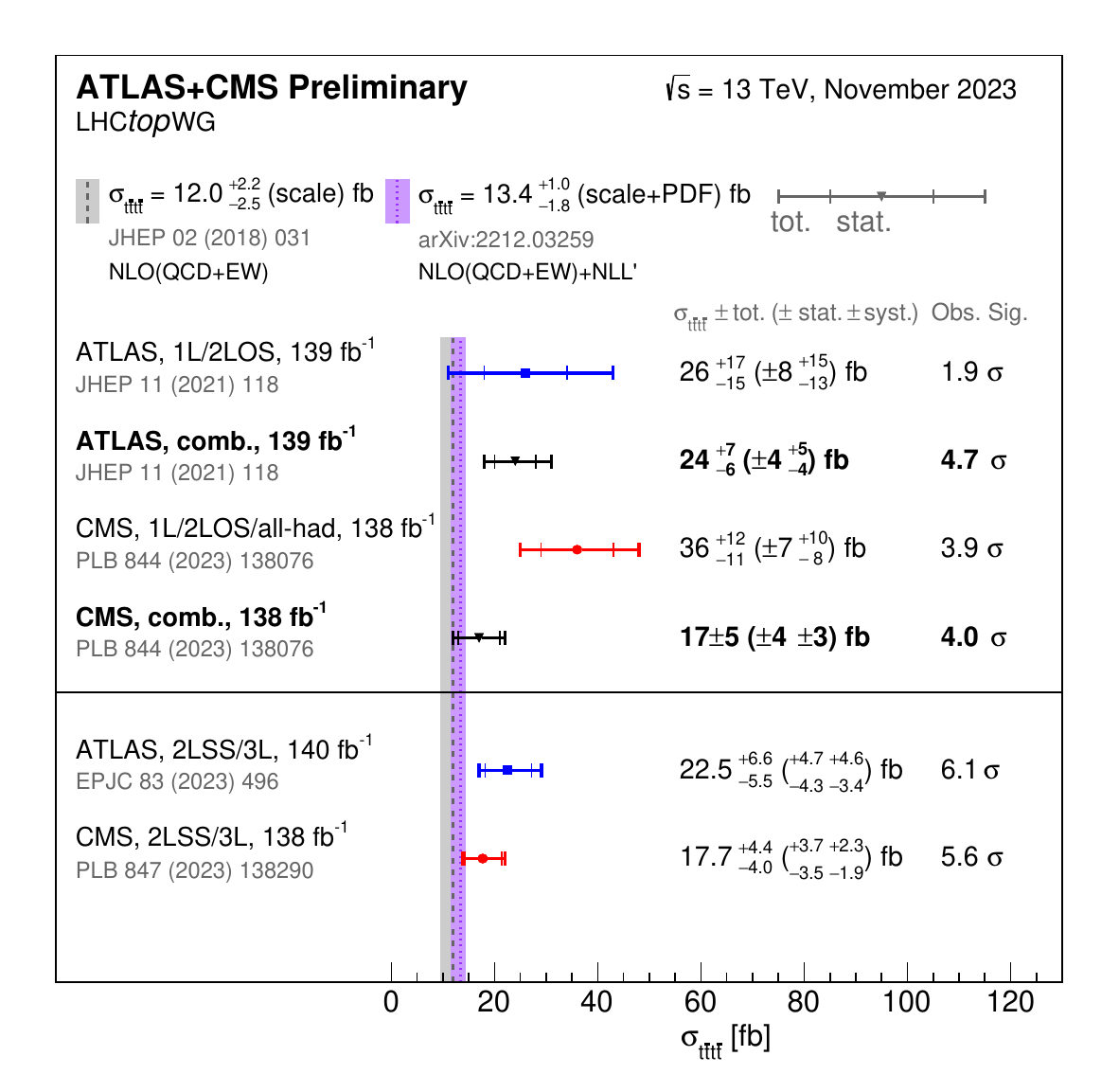}
    \caption{Summary of ATLAS and CMS measurements of the $t\bar{t}t\bar{t}$ production cross section at 13 TeV in various channels~\cite{toplhcwg}.}
    \label{fig:4top}  
\end{figure}

The 4-top ($t\bar{t}t\bar{t}$) production is a very rare process sensitive to Higgs boson properties and beyond SM particles. The $t\bar{t}t\bar{t}$ production process is observed by both collaborations, CMS~\cite{CMS:2023ftu} and ATLAS~\cite{ATLAS:2023ajo}, using proton-proton collisions at 13~TeV with the full run 2 data. 
The measurements are based on multilepton final states, two same-sign, three and four leptons plus additional jets. The main background contributions are processes of top-quark pairs associated with a W, Z or Higgs boson (known as $t\bar{t}X$). The leading sources of systematic uncertainties are the signal MC generator choice and parton shower modelling, the $t\bar{t}X$ background theory modelling, and the jet and b-jet related sources of uncertainty. The results of the 4-top production cross sections are shown in Fig.~\ref{fig:4top}.

\section{\boldmath Inclusive and differential $t\bar{t}W$ cross sections}
\begin{figure}[h]
    \centering
    \includegraphics[width=0.75\linewidth]{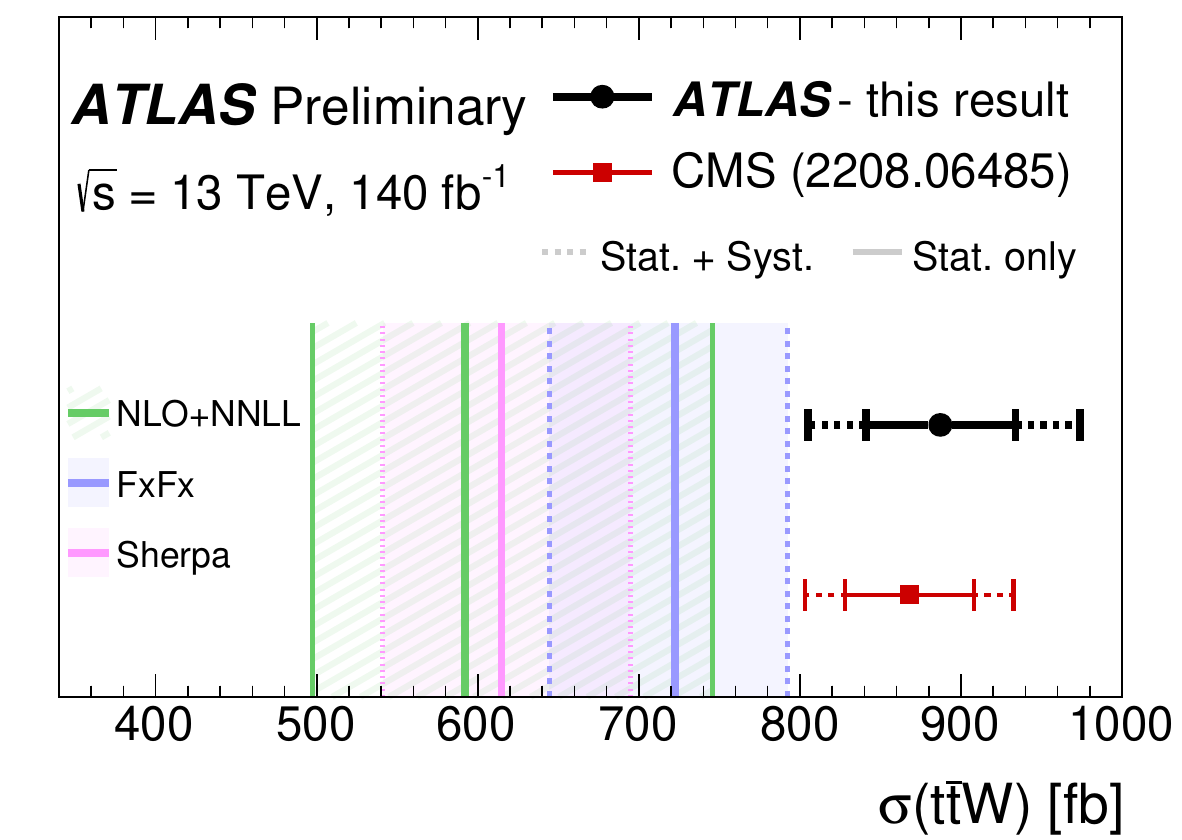}
    \caption{Comparison of the measured inclusive $t\bar{t}W$ cross sections to the theoretical predictions from Sherpa, the MGNLOPY FxFx prescription including electroweak corrections and the NLO+NNLL prediction~\cite{ATLASttW}.}
    \label{fig:ttW}  
\end{figure}

Measurements of both the inclusive production cross sections of a top-antitop quark pair in association with a W boson ($t\bar{t}W$) by the CMS~\cite{CMS:2022tkv} and ATLAS~\cite{ATLASttW} Collaborations and the differential production cross sections by the ATLAS~\cite{ATLASttW} Collaboration are presented.  The measurements are performed using the full run 2 data targeting final states with two same-sign or three isolated electrons and muons. The preliminary ATLAS results are superseded in Ref.~\cite{ttWarxiv}. The $t\bar{t}W$ cross sections have been consistently measured above expectation. These results, shown in Fig.~\ref{fig:ttW}, are compatible at the 1.5 and 2 $\sigma$ level for ATLAS and CMS, respectively. The leading uncertainties are coming from signal modelling and background normalizations mainly $t\bar{t}Z$, $WZ$ and misidentified leptons.

ATLAS absolute and normalised differential cross section measurements characterise this process in detail for the first time. Several particle-level observables are compared to theoretical predictions from different MC generators which are in good agreement with the normalised differential cross section results. 

\section{\boldmath Inclusive and differential $t\bar{t}b\bar{b}$  cross sections}

\begin{figure}[h]
    \centering
    \includegraphics[width=0.95\linewidth]{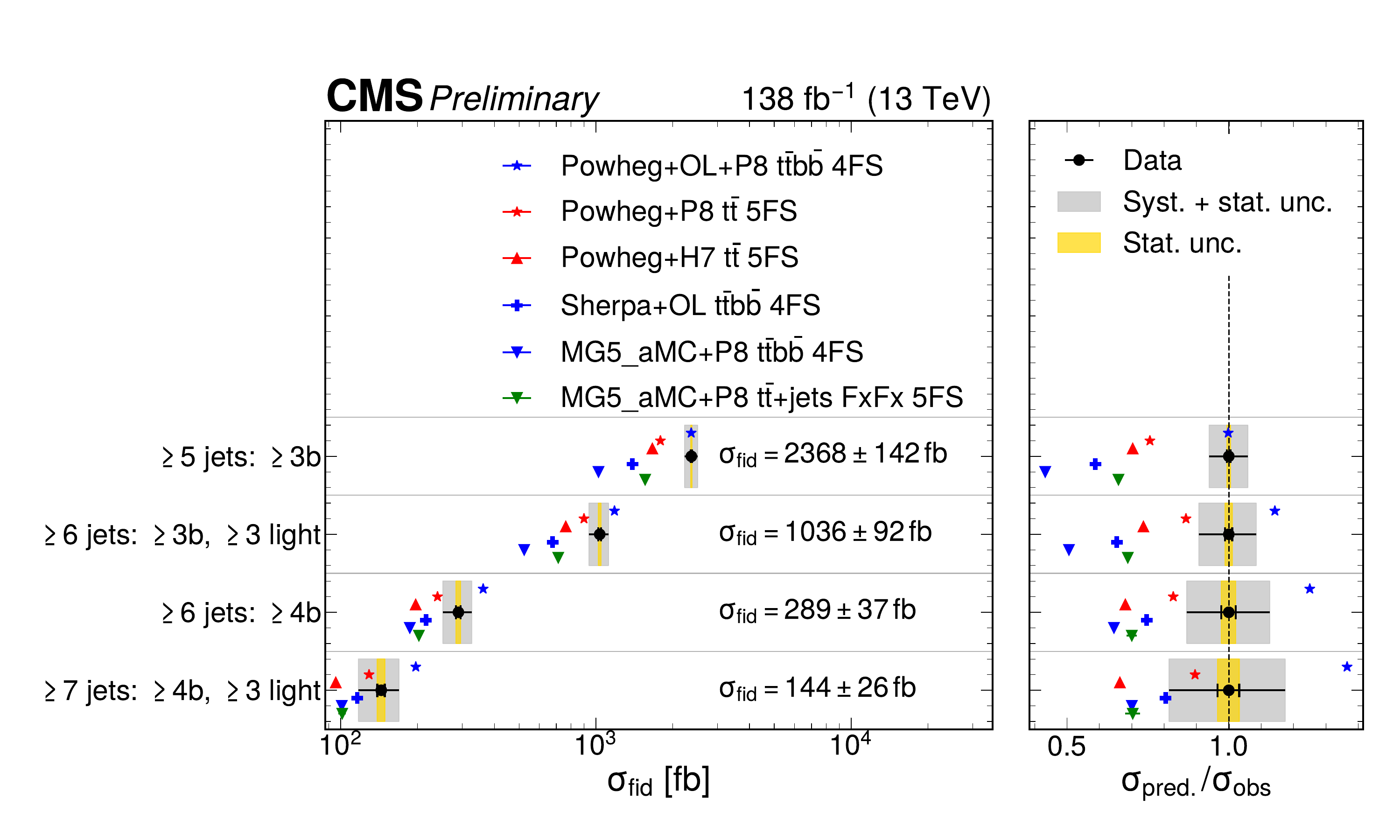}
    \caption{Measured inclusive cross sections for each considered phase space, compared to predictions from different $t\bar{t}b\bar{b}$ simulation approaches shown as coloured symbols~\cite{CMSttbb}.} %The blue colour is reserved for models using massive b quarks and NLO QCD $t\bar{t}b\bar{b}$ matrix elements, while red is used for the inclusive $t\bar{t}$ generators at NLO in QCD with massless b quarks. The right panel shows the ratios between the predicted and measured cross sections.}
    \label{fig:ttbb}  
\end{figure}

The measurement by the CMS Collaboration presents the most precise inclusive production cross sections of $t\bar{t}b\bar{b}$ up to date~\cite{CMSttbb}  using proton–proton collisions at a centre-of-mass-energy of 13 TeV and an integrated luminosity of 138~fb$^{-1}$. The measurements are performed in the lepton+jets decay channel of the top-quark pair, using events containing exactly one isolated lepton (muon or electron) and at least five jets. The cross sections are measured in four fiducial phase space regions requiring different jet, b-jet, and light-jet multiplicities, targeting different aspects of the $t\bar{t}b\bar{b}$ process. The uncertainties are dominated by systematic sources with the leading sources originating from the calibration of the b tagging and of the jet energy scale, and from the choice of renormalization scale in the signal $t\bar{t}b\bar{b}$ and background $t\bar{t}$ processes. The considered event generators predictions are 10-50\% lower than the measured values in the different phase space regions as shown in Fig.~\ref{fig:ttbb}.

Differential cross section measurements are performed as a function of several
observables compared with predictions from several event generators. The normalized differential cross sections show varying degrees of compatibility with the theoretical predictions, however it is found that none generator simultaneously describes all the measured distributions. These preliminary results are superseded in the paper publication~\cite{CMS:2023xjh}.
 
\section{\boldmath Charge asymmetry in $t\bar{t}\gamma$}
\begin{figure}[h]
    \centering
    \includegraphics[width=0.9\linewidth]{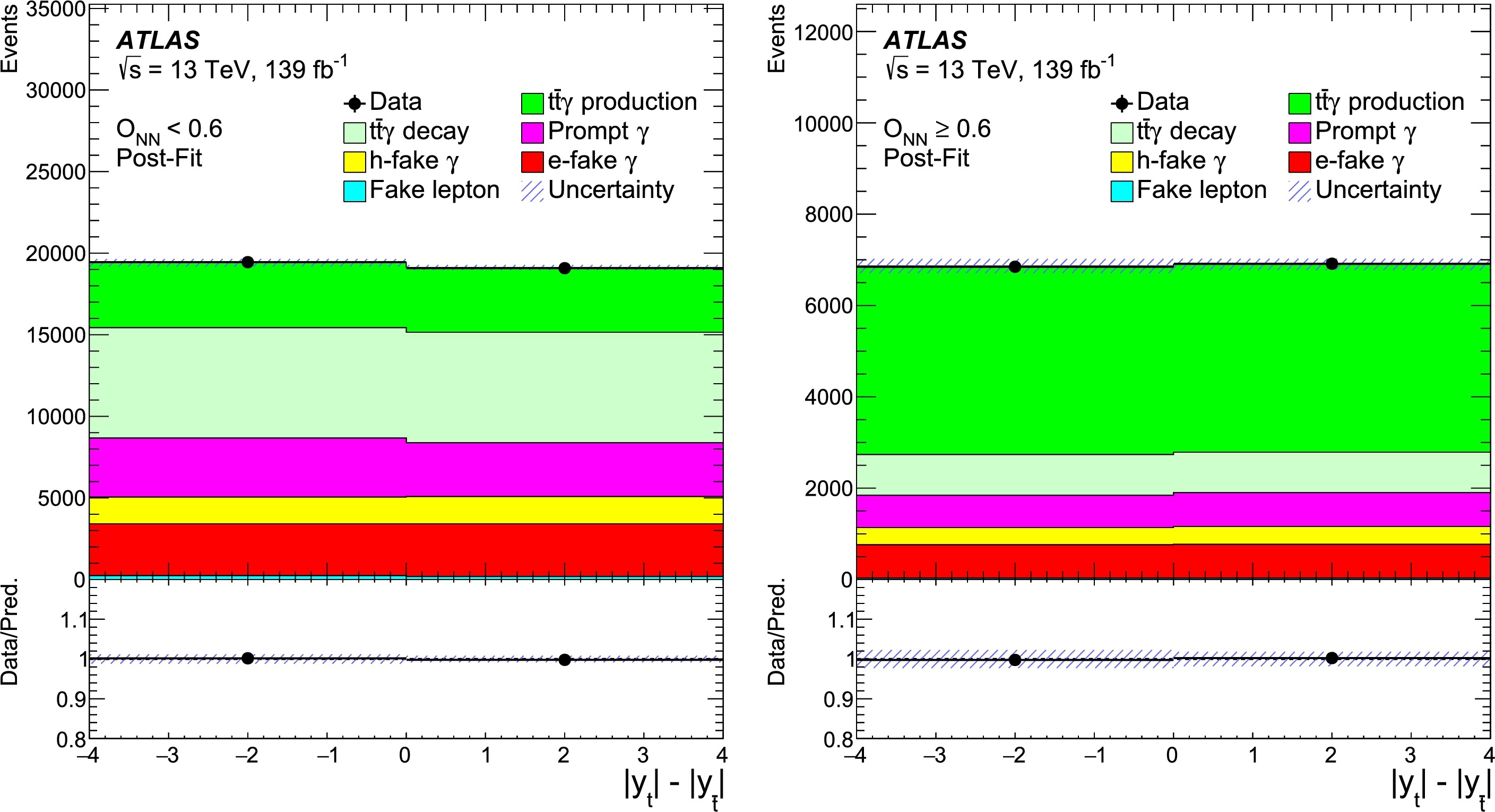}
    \caption{The distributions of $|y_t|-|y_{\bar{t}}|$ after the fit in the two regions defined by the NN output~\cite{ATLAS:2022wec}.}
    \label{fig:ttgamma}  
\end{figure}

A measurement of the top-quark pair charge asymmetry in $t\bar{t}\gamma$ is presented~\cite{ATLAS:2022wec}. It is performed using proton–proton collision data collected with the ATLAS detector at a centre-of-mass-energy of 13 TeV during the years 2015–2018, corresponding to an integrated luminosity of 139~fb$^{-1}$. The selected events have exactly one photon, one lepton, and at least four jets, of which at least one is b-tagged. The separation between signal ($t\bar{t}\gamma$  production where the $\gamma$ comes only from $t\bar{t}$ production) and background is enhanced using a Neural Network (NN) approach. The charge asymmetry is obtained from the distribution of the difference of the absolute rapidities of the top quark and antiquark ($|y_t|-|y_{\bar{t}}|$), as shown in Fig.~\ref{fig:ttgamma}, using a profile likelihood unfolding approach. It is measured to be A$_C$ = -0.003 $\pm$ 0.024(stat) $\pm$ 0.017(syst) in agreement with the standard model expectation.
%---------------------------------------------------------------------------------------%

\section{Summary}
Low cross section top quark processes like $t\bar{t}X$ and $t\bar{t}t\bar{t}$ processes are already accessible at the LHC given the large amount of data recorded during run 2. Inclusive and differential cross sections are measured by the CMS and ATLAS Collaborations probing the top quark properties. All results described in these proceedings are in good agreement with SM predictions within the limited systematic uncertainties.

\bibliographystyle{amsplain}

\end{document}